\documentclass[12pt,preprint]{aastex}
\usepackage{url}

\shorttitle{Dynamics of Magnetized Vortex Tubes in the Solar Chromosphere}

\title{Dynamics of Magnetized Vortex Tubes\\in the Solar Chromosphere}

\author{I.~N. Kitiashvili$^{1,2}$, A.~G. Kosovichev$^1$, N.~N. Mansour$^3$, A.~A. Wray$^3$}
\affil{$^1$Stanford University, Stanford, CA 94305, USA}
\altaffiltext{1}{e-mail: irinasun@stanford.edu}
\affil{$^2$ Kazan Federal University, Kazan, 420008, Russia}
\affil{$^3$NASA Ames Research Center, Moffett Field, Mountain View, CA 94040, USA}

\begin{document}
\begin{abstract}
We use 3D radiative MHD simulations to investigate the formation and dynamics of small-scale (less than 0.5~Mm in diameter) vortex tubes spontaneously generated by turbulent convection in quiet-Sun regions with initially weak mean magnetic fields. The results show that the vortex tubes penetrate into the chromosphere and substantially affect the structure and dynamics of the solar atmosphere. The vortex tubes are mostly concentrated in intergranular lanes and are characterized by strong (near sonic) downflows and swirling motions that capture and twist magnetic field lines, forming magnetic flux tubes that expand with height and which attain magnetic field strengths ranging from 200~G in the chromosphere to more than 1~kG in the photosphere. We investigate in detail the physical properties of these vortex tubes, including thermodynamic properties, flow dynamics, and kinetic and current helicities, and conclude that magnetized vortex tubes provide an important path for energy and momentum transfer from the convection zone into the chromosphere.
\end{abstract}

\section{Introduction}
Interest in vortex tube dynamics of the quiet Sun was recently initiated by the detection of ubiquitous small-scale swirling motions in the  photosphere \citep{wang1995,potzi2005,bonet08,bonet10,Balmaceda2010,steiner2010} and the chromosphere \citep{Wedemeyer-Bohm2009} with high-resolution solar telescopes. Previous to this discovery, vortex tubes on the Sun were predicted by theoretical models \citep[e.g.,][]{Stenflo1975} and numerical simulations \citep[e.g.,][]{brand1996,stein2000}, giving a clear illustration of the turbulent nature of solar convection. Both observations and numerical simulations show concentrations of vortex tubes in the intergranular lanes. According to recent radiative hydrodynamic simulations, vortical motions can be also form inside granules \citep{kiti2012}. These simulations have also shown that vortex tube formation in the near-surface layers can be caused by two basic mechanisms associated with: 1) small-scale convective instability developing inside granules, and 2) the Kelvin-Helmholtz instability of shearing flows.

The convective instability leads to formation of a vortex sheet and its subsequent overturning during a  localized upflow (plume) or splitting of a granule. The process of the vortex sheet overturning, which results in a vortex tube, is often accompanied by a gradual migration of the vortex tube into an intergranular lane \citep[see Fig.~2 in][]{kiti2012}. Shearing flows that lead to the development of the Kelvin-Helmholtz instability can be present in both granules and intergranular lanes. However, in the intergranular lanes the shearing flows are stronger and can lead to a series of vortices (resembling the Karman vortex street). Also, converging downflows in the intergranular lanes make the vortex tubes more stable, with characteristic lifetimes up to 40~min, whereas inside granules the lifetime is less than 10~min. These processes can explain why the observed vortex tubes are predominantly concentrated in the intergranular lanes.

Numerical simulations also show connections between vortex tube dynamics and various other solar phenomena, such as the hydromagnetic dynamo \citep{brand1996}, spontaneous organization of emerged magnetic field into self-maintained pore-like structures \citep{kiti2010b}, excitation of acoustic waves in the quiet Sun \citep{kiti2011}, and others. In this Letter, we present new  numerical simulations that demonstrate important links between the turbulent subsurface layers and the solar atmosphere though the dynamics of penetrating vortex tubes.

\section{Computational setup}
Numerical simulations of the quiet Sun are performed by using a 3D radiative MHD code (`SolarBox') developed at the NASA/Ames Research Center and the Stanford Center for Turbulence Research by Alan Wray and his colleagues \citep{jacoutot08a} for modeling the outer part of the solar convection zone and lower atmosphere in a cartesian geometry. The code was developed for realistic-type numerical simulations of the Sun pioneered by \cite{nordlund2001} and uses a tabular real-gas equation of stat.  Radiative energy transfer is calculated with a 3D multi-spectral-bin method between fluid elements, assuming local thermodynamic equilibrium and using the OPAL opacity tables \citep{Rogers1996}.  Initialization is done from a standard model of the solar interior \citep{chris1996}.

The physical description of the dynamical properties of solar convection was improved through the implementation of subgrid-scale turbulence models, which effectively increase the Reynolds number and allow better resolution of essential turbulent scales. This approach, based on Large-Eddy Simulation (LES) models of subgrid turbulence, has demonstrated good agreement of numerically modeled acoustic wave excitation with observations \citep{jacoutot08a} and has helped improve understanding of wave excitation mechanisms \citep{kiti2011}, formation of magnetic structures \citep{kiti2010b}, and Evershed flows in sunspots \citep{kiti2009}. The simulations in this paper were obtained using a Smagorinsky eddy-viscosity model \citep{Smagorinsky1963} in which the compressible Reynolds stresses were calculated in the form \citep{Moin1991, jacoutot08a}: $\tau_{ij}=-2C_S\triangle^2|S|(S_{i,j}-u_{k,k}\delta_{ij}/3)+2C_C\triangle^2|S|^2\delta_{ij}/3$, where the Smagorinsky coefficients  $C_S=C_C=0.001$, $S_{ij}$ is the large-scale stress tensor, and $\triangle\equiv(dx\times dy\times dz)^{1/3}$ with $dx$, $dy$ and $dz$ being the grid-cell dimensions.

In the current study, the simulation results were obtained for a computational domain of $6.4\times6.4\times6.2$~Mm$^3$, including a 1~Mm high layer of the atmosphere, with a grid spacing of $dx=dy=12.5$~km and $dz=10$~km. The lateral boundary conditions are periodic. The top boundary is open to mass, momentum, and energy transfers and also to radiative flux. The bottom boundary is open for radiation and flows, and simulates energy input from the interior of the Sun. We focus mostly on a case with an initially uniform vertical magnetic field, $B_z=10$~G, representing quiet-Sun conditions (far from sunspots and active regions).

\section{Formation of vortex tubes by turbulent convection}

Vortex tubes are formed by turbulent convection in near-surface layers of the convective zone \citep[e.g.][]{stein2000,kiti2012}. The vortex tubes represent compact low-density structures up to 0.5~Mm in diameter and with high-speed swirling motion reaching up to 12~km/s. The vortex cores are characterized by strong downflows (up to 8~km/s) and lower temperature. Large vortex tubes can extend deeper than 300~km below the surface.

\begin{figure}
\begin{center}
\includegraphics[width=1\linewidth]{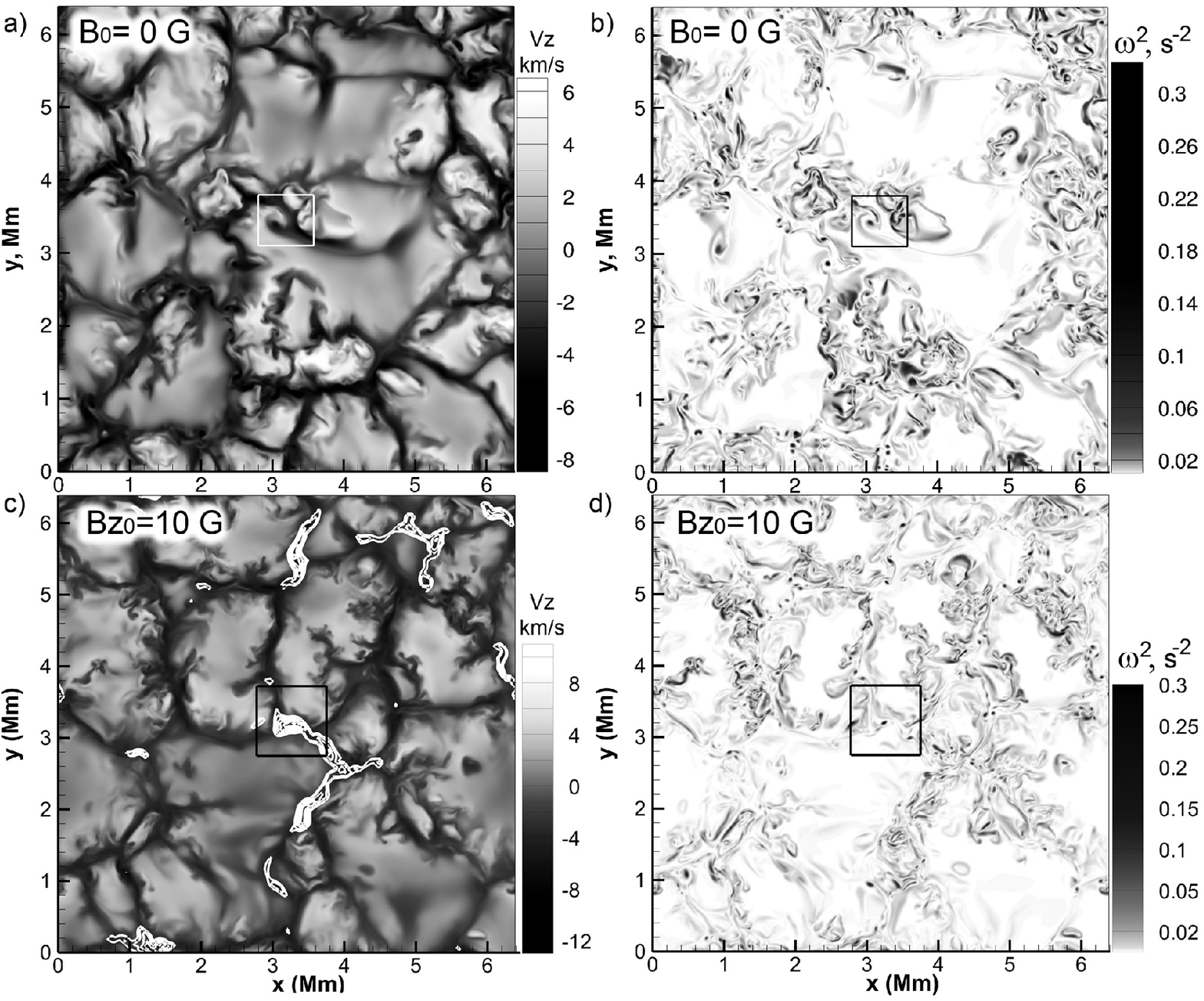}
\end{center}
\caption{Snapshots of vertical velocity V$_{\rm z}$ and enstrophy $\omega^2=(\nabla\times\vec u)^2$ on the solar surface for two simulation cases: without magnetic field (panels {\it a} and {\it b}) and with initial uniform weak vertical magnetic field, $Bz_0=10$~G (panels {\it c} and {\it d}). Examples of small-scale vortex tubes are indicated by white and black squares. \label{fig:mag} }
\end{figure}

Our previous simulations \citep{kiti2012} revealed two basic mechanisms of vortex tube formation: one due to a granular instability (vortex sheet overturning) and another due to the Kelvin-Helmholtz instability in shearing flows. Vortex tubes can form in intergranular lanes and in granules but are mostly concentrated in the intergranular lanes (Fig.~\ref{fig:mag}{\it a, b}). These physical mechanisms of vortex tube formation are purely hydrodynamic, but in the real Sun vortices are expected to strongly interact with ubiquitous magnetic fields. However, neither simulations nor observations have shown a clear correlation between vortex motions and magnetic field concentrations, that is, not every vortex is accompanied by a strong magnetic field concentration. This fact has also been shown in simulations using a shallow domain \citep[1.4~Mm in total height;][]{sh2011,Moll2011}.

In weak magnetic field regions, magnetic patches follow convective motions. Concentration and magnification of magnetic field by swirling motions can stabilize the vortex tube structure and decrease the influence of surrounding turbulent flows. In our simulation case, we introduce a 10~G, initially uniform, vertical magnetic field. This field gets quickly concentrated, mostly in intergranular lanes, and we find that the strongest magnetic field ($\sim 1$~kG) concentrations are often associated with vortices (Fig.~\ref{fig:mag}{\it c, d}).

\section{Dynamics and properties of vortex tubes in the chromosphere}

\begin{figure}
\begin{center}
\includegraphics[width=1\linewidth]{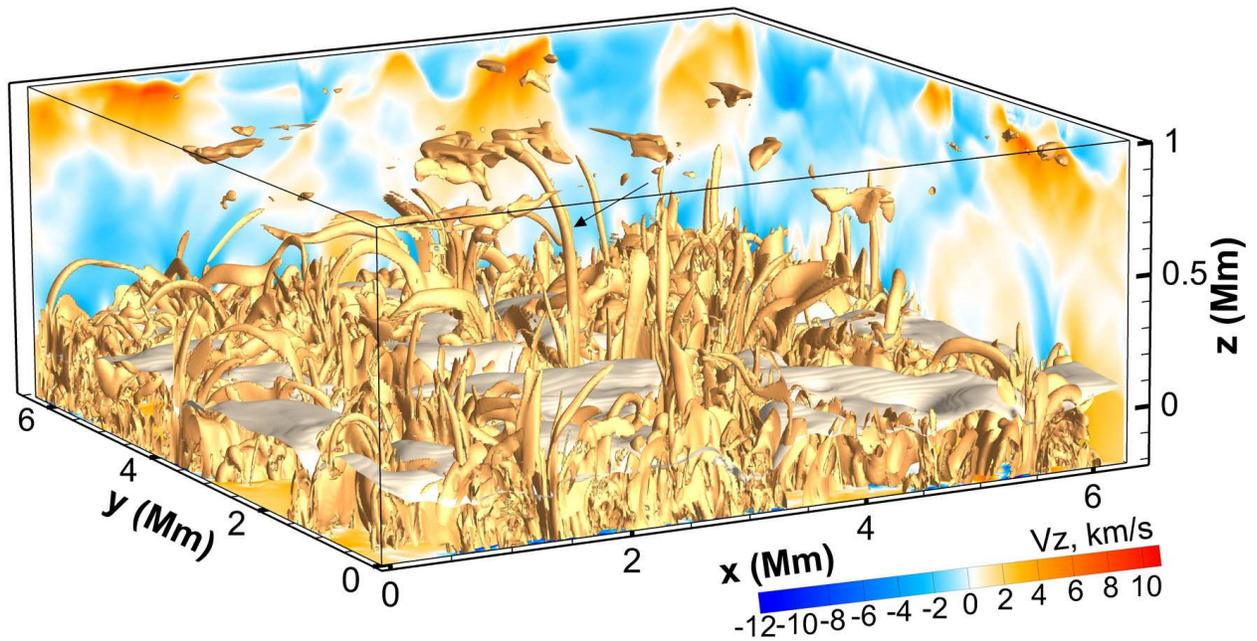}
\end{center}
\caption{3D snapshot of the top part of the computational domain shows the penetration of convective vortex tubes (yellow isosurfaces) from the subphotosphere into the low and mid chromospheric layers. The horizontal wavy surface indicates the distribution of temperature at 6400~K, and corresponds to a photosphere layer. The vortex tubes (yellow isosurfaces) are shown for the enstrophy value of 0.0075 s$^{-2}$. The vertical slices in the back illustrate the vertical velocity distribution. Blue color corresponds to downflows, yellow-red shows upflows. \label{fig:total}}
\end{figure}

A new interesting result of our simulations is the extension of turbulent vortex tubes from the convection zone into the convectively stable atmospheric layers. Figure~\ref{fig:total} illustrates a snapshot of enstrophy distribution showing vortex tube structures (yellow isosurfaces) above the photosphere (the horizontal wavy light surface shows the 6400~K near-surface layer). These vortex tubes are mostly concentrated in the intergranular lanes and often form arc-shaped structures above the surface. Other vortices penetrate almost vertically into the higher chromospheric layers (an example of such an extended vortex tube is indicated by the arrow; we will consider its structure in detail below). Local upflows (red color on vertical slices, Figure~\ref{fig:total}) cause stretching of the vortex arcs, and nearby vortices can destroy them. Finally, propagating shock waves interact with the vortex tubes in the higher chromospheric layers. The overall chromospheric dynamics driven by turbulent convection is thus very complicated. The effect of vortex penetration into the chromosphere is mostly hydrodynamic, as observed in simulations with and without magnetic field. However, the magnetic field tends to be captured and concentrated in the vortex tubes, causing new dynamical effects.

\begin{figure}
\begin{center}
\includegraphics[width=1\linewidth]{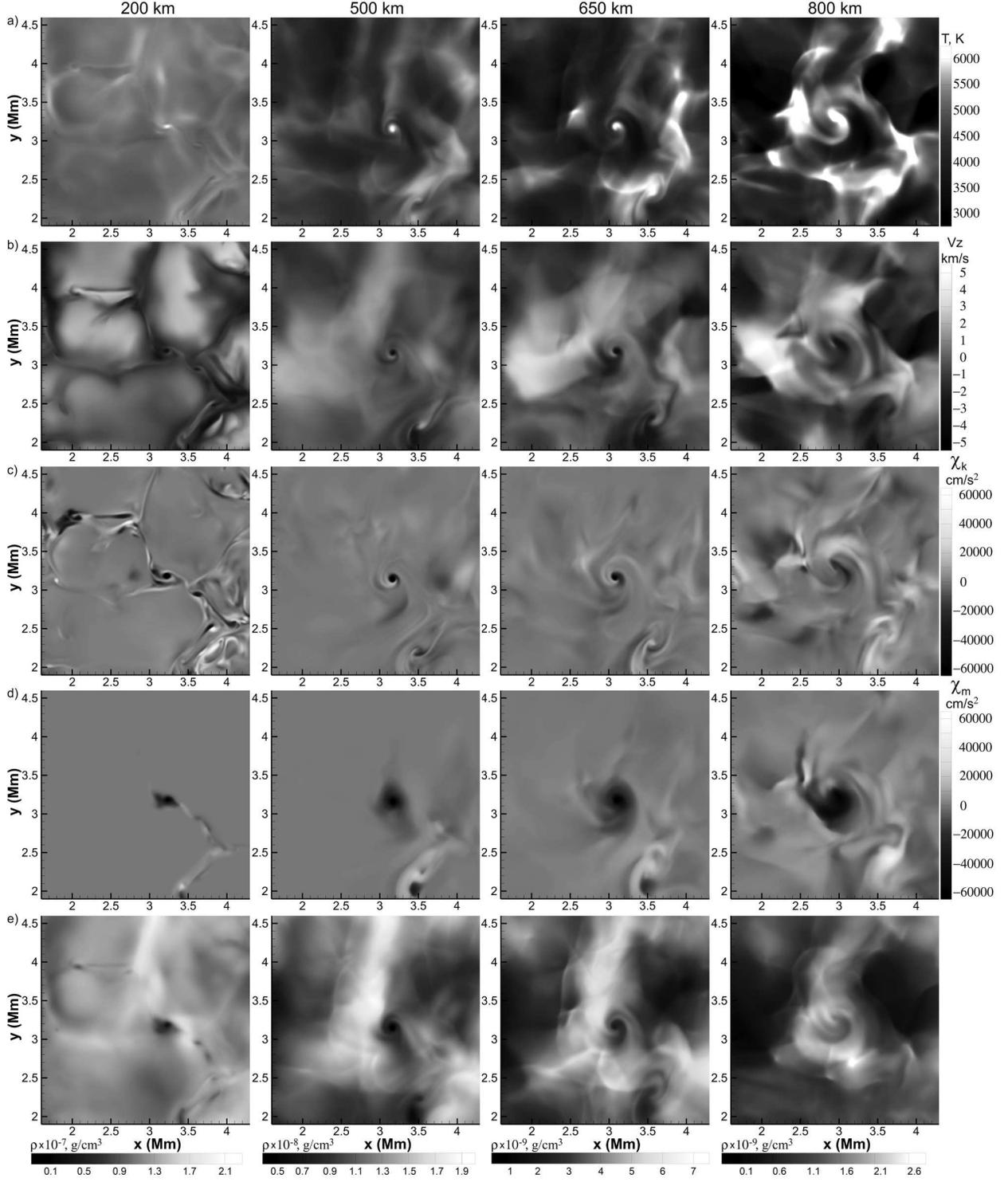}
\end{center}
\caption{Vortex tube structure in four different atmospheric layers: 200~km, 500~km, 650~km and 800~km above the solar surface (from left to right). Each row corresponds to various quantities: {\it a}) temperature T, {\it b}) vertical velocity Vz, {\it c}) kinetic helicity $\chi_k$, {\it d}) current helicity $\chi_m$, and {\it e}) density $\rho$. \label{fig:snH}}
\end{figure}

The structure of the vortex tube indicated by the arrow in Figure~\ref{fig:total} is illustrated in Figure~\ref{fig:snH} at different heights: 200~km, 500~km, 650~km and 800~km above the surface. The temperature distribution  (Fig.~\ref{fig:snH}, row {\it a}) shows local heating of the vortex core region, whereas in the subphotospheric layers the core vortex temperature is lower than in the surrounding plasma. For this moment of time, the swirling motions in the vortex region are characterized mostly by highly turbulent downflows  (Fig.~\ref{fig:snH}{\it b}), but some upflows are noticeable near the edge of the vortex tube, the size of which is expanding with height. The current helicity, calculated in Alfv\'en units as $\chi_m=\frac{1}{4\pi\rho}\vec{B}\cdot (\nabla\times\vec{B})$, forms a sheet-like structure oriented along the intergranular lane near the photospheric layers (Fig.~\ref{fig:snH}{\it d}). The current sheet structure gradually changes orientation in the higher layers and becomes more circular (Fig.~\ref{fig:snH}{\it d}). The current helicity structure is more diffuse than the kinetic helicity, $\chi_m=\vec{u}\cdot (\nabla\times\vec{u})$, shown in Figure~\ref{fig:snH}{\it c}. The density distribution in the lower atmosphere is similar to the surface layers, but the vortex tube structure becomes more complicated with height, forming a ring-like structure at $\sim800~$km above the surface (Fig.~\ref{fig:snH}{\it e}).

\begin{figure}
\begin{center}
\includegraphics[width=1\linewidth]{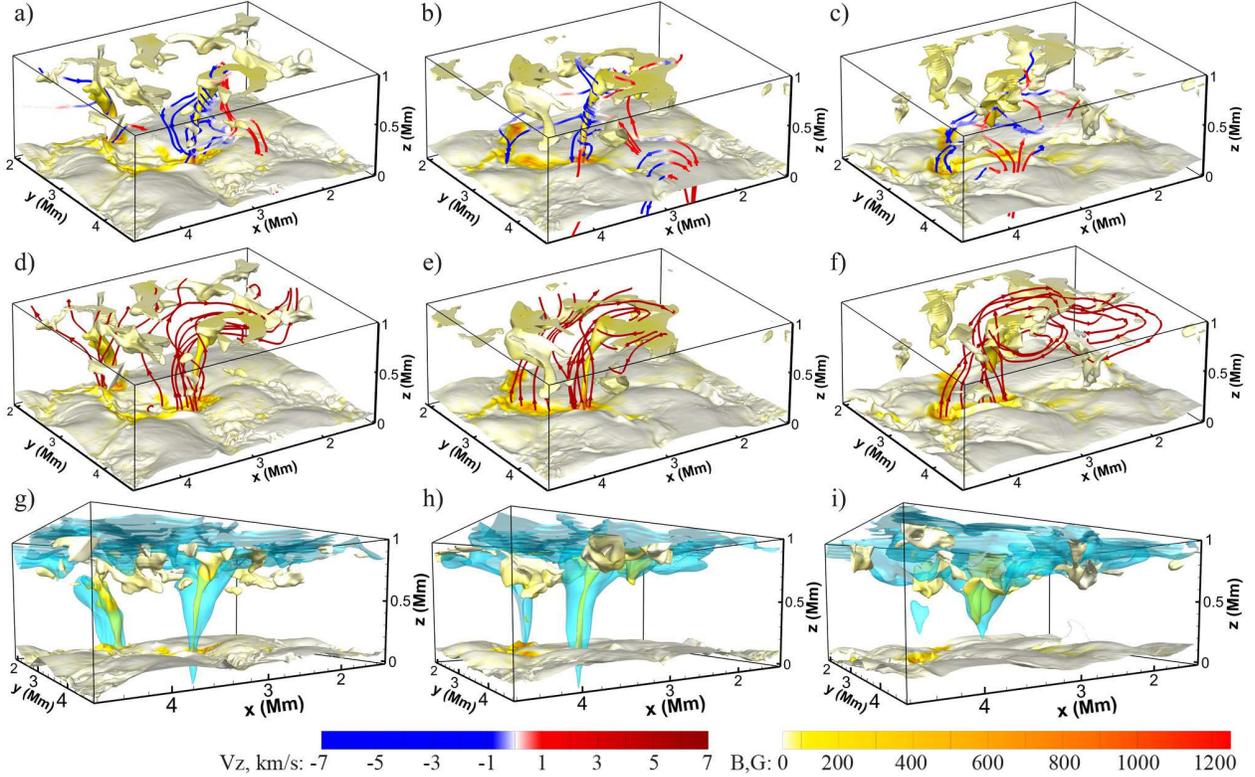}
\end{center}
\caption{Evolution of the velocity field (streamlines in panels {\it a-c}), magnetic topology (streamlines in panels {\it d-e}), and plasma parameter $\beta$ (blue isosurface for $\beta=3$ in panels {\it g-i}). Each column corresponds to simulation data 3 min apart. The grey isosurface shows T=5800~K; additional coloring from light yellow to orange indicates variations of the magnetic field strength in the range from 0 to 1200~G. Coloring of the velocity streamlines in panels {\it a-c} corresponds to vertical velocities from $-7$~km/s (blue) to $+7$~km/s (red). \label{fig:spaps} }
\end{figure}

\begin{figure}
\begin{center}
\includegraphics[width=0.9\linewidth]{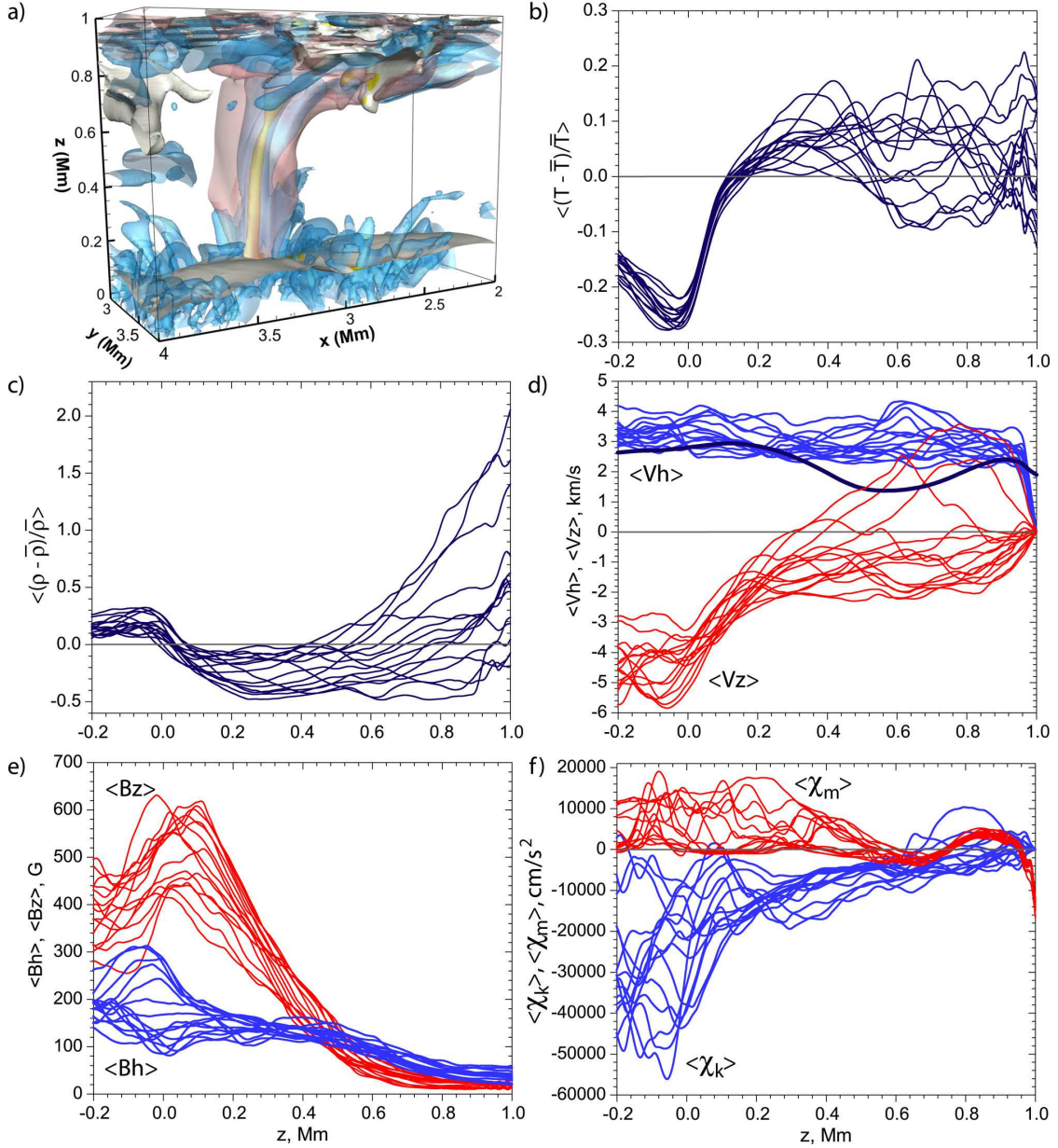}
\end{center}
\caption{{\it a}) 3D rendering of a vortex tube indicated by the arrow in Figure~\ref{fig:total} shows the relative distribution of the kinetic (blue isosurface) and current helicities (pink) for value $-5000$~cm/s$^2$. The grey-yellow isosurface shows the distribution of temperature T=5800~K. Panels {\it b-f} show the distribution with height of mean vortex tube properties at different moments of time with 20~sec cadence: {\it b}) relative temperature variations, {\it c}) relative density,  {\it d}) horizontal speed $\langle$Vh$\rangle$ and vertical velocity $\langle$Vz$\rangle$, {\it e}) horizontal $\langle$Bh$\rangle$ and vertical $\langle$Bz$\rangle$ magnetic field strength, {\it f}) kinetic $\langle\chi_k\rangle$ and magnetic $\langle\chi_m\rangle$ helicities. Thick blue solid curve in panel {\it d} shows the mean horizontal speed averaged over the whole domain and time. The systematic variations at z $> 0.95$~Mm are due to the top boundary conditions. \label{fig:profiles}}
\end{figure}

Figure~\ref{fig:spaps} shows the time evolution of the velocity streamlines (panels {\it a-c}), magnetic field lines (panels {\it d-f}), and the ratio of gas pressure to magnetic pressure (plasma $\beta$) for three moments separated by  3~min. The structure of the vortex tube in the middle column is shown in more detail in Fig.~\ref{fig:profiles}{\it a}. In Figure~\ref{fig:spaps}, the grey-yellow isosurface corresponds to a temperature of 5800~K. Color patches on this surface indicate variations of  magnetic field strength as indicated in the right color bar. The strongest magnetic field concentrations ($\sim 1.2$~kG) are associated with the vortex tubes in the photospheric layer; and the field strength decreases to $\sim 200$~G in the upper layers of our domain ($\sim 1$~Mm above the photosphere).

The numerical simulations show the penetration and dynamics of the vortex tube into the chromosphere. The vortex core contains very compact helical downflows, and we observe that the vortex pulls granular fluid upward which then reverses into the downflows (Fig.~\ref{fig:spaps}{\it a}). The magnetic field at this stage of vortex evolution continues to concentrate in the vicinity of the vortex by following the swirling turbulent motions (Fig.~\ref{fig:spaps}{\it d}). These strong helical flows capture and twist the magnetic field lines. Also, the helical magnetic loops formed by vortex tubes have a tendency to move upward due to local upflows near the vortex core.
Three minutes later, the helical downflows have become more compact and stronger (Fig.~\ref{fig:spaps}{\it b}), but the vortex is affecting a larger surrounding area. We begin to see evidence of vortex decay when this vortex starts interacting with others by sharing with them a part of the downflow (Fig.~\ref{fig:spaps}{\it b}). Finally, during the next three minutes, the photospheric and chromospheric parts of the vortex tube become disconnected but still continue to evolve. Figure~\ref{fig:spaps}{\it c} shows remnants of the initially strong helical flows in the atmosphere. At this moment, they are still weakly helical and become captured by another growing vortex. The magnetic field lines also keep their helical topology and start to diffuse (Fig.~\ref{fig:spaps}). We show the plasma parameter $\beta\equiv\frac{8\pi p}{B^2}=3$ level as blue isosurfaces in Figures~\ref{fig:spaps}{\it g-i}; the value $\beta=1$ is reached only in a small region of the vortex core. This parameter shows that magnetic effects play a significant role in the photospheric layers of the vortex tube and that the region of their influence rapidly expands with height. At the decay stage of the vortex tube (Fig.~\ref{fig:spaps}{\it i}), magnetic effects are significant only in the upper layers.

The relative role of kinematic and magnetic effects of the swirling motions is illustrated by the kinetic, $\chi_k$, and magnetic, $\chi_m$, helicities. An example of relative distribution for both helicities is shown in Figure~\ref{fig:profiles}{\it a} (the kinetic helicity is in blue, and the current one is in pink) for the vortex tube that is indicated by the arrow in Figure~\ref{fig:total}. Blue and pink isosurfaces correspond to helicity values of $-5000$~cm/s$^2$; the current helicity is calculated in Alfv\'en units and has the same dimension as the kinetic helicity. In Figure~\ref{fig:profiles}a, we also plot the temperature isosurface for 5800~K, which has a very compact structure of a complicated chiralical shape, expanding into the higher layers of the atmosphere. The distribution of the kinetic helicity is more compact than the current helicity, meaning that the swirling flows in the vortex tube are more compact than the twisted magnetic field lines.

In general, the dynamics of the subsurface and near-surface layers is dominated by turbulent convective motions, while magnetic effects are noticeable in the small-scale magnetic flux concentrations in the intergranular lanes (magnetic flux tubes). In higher atmospheric layers, magnetic effects are stronger because of the fast decrease of gas pressure, which leads to expansion of the magnetic flux tubes.

To investigate the properties of the magnetized vortex tube with height, we selected a region inside the $-5000$~cm/s$^2$ isosurface of the current helicity.  In this region we have plotted the mean values of temperature, vertical and horizontal velocities, magnetic field, and kinetic and current helicities as a function of height for different moments of time with a cadence of 20 sec (Fig.~\ref{fig:profiles}{\it b-f}). The values of temperature and density are shown as perturbations from and normalized by the mean values: $(T_{vortex}-\overline{T})/\overline{T}$ and $(\rho_{vortex}-\overline{\rho})/\overline{\rho}$. The temperature distribution shows a deficit in the convective layers of the vortex tube. Above the photosphere, the temperature in the vortex tube increases, and we can see heating in the vortex core (Fig.~\ref{fig:profiles}{\it b}). Occasional temperature decreases above $\sim 500$~km reflect the dynamically oscillatory behavior of the vortex tube. The density distribution (Fig.~\ref{fig:profiles}) shows an increase below the surface due to mass concentration around the vortex core, which has significantly lower density. Above the surface, the mean density perturbation in the tube first decreases and then increases above 200~km (Fig.~\ref{fig:profiles}{\it c}).

The vertical distribution of the mean velocity inside the vortex tube shows very different properties for the horizontal and vertical components. The mean horizontal speed (blue curves, Fig~\ref{fig:profiles}{\it d}) is almost constant along the vortex tube, with relatively small fluctuations in time around the mean speed of $\sim 3.5$~km/s. It is interesting that the mean horizontal speed, averaged over time and over the whole domain, (thick dark blue curve) shows a decrease at $400-800$~km above the surface, but inside the vortex tube there is no such decrease. In contrast to the horizontal speed, the vertical velocity component (red curves, Fig,~\ref{fig:profiles}{\it d}) is very dynamic and is characterized by predominant downflows; however, local upflows can be detected inside the vortex tube.

The vertical component of magnetic field is significantly stronger than the magnitude of the horizontal field (Fig.~\ref{fig:profiles}{\it e}). Both vertical (red curves) and horizontal (blue curves) fields show a similar tendency to decrease in the atmospheric layers, which is reflected in the expanding topology of the flux tube. At a height of about 500~km, the magnitude of the vertical component of magnetic field is smaller than the horizontal component because the magnetic field lines become more twisted by the vortex. The mean kinetic helicity (blue curves, Fig.~\ref{fig:profiles}{\it f}) is significantly greater than the mean current helicity (red curves) because swirling motions in the tube are accompanied by strong downflows, while the strongest magnetic field is only weakly twisted.

\section{Conclusion}
The formation and dynamics of small-scale vortex tubes play key roles in various processes in solar surface convection and in the solar atmosphere. Our radiative MHD simulations reveal vortex tubes formed by turbulent convection penetrating from the subphotosphere into the chromosphere. These vortex tubes cause significant qualitative changes in atmospheric dynamics, leading to strong variations in the thermodynamic structure through local heating and density variations, generating twisted magnetic flux tubes, and creating local twisted upflows into the chromosphere. Strong localized swirling motions occupy large areas around the vortex tubes, capturing and twisting magnetic field lines from nearby magnetic structures. As a result of these phenomena, magnetized vortex tubes generated by turbulent convective motions provide a very important link for energy and momentum exchange between the surface layers and the chromosphere.

\acknowledgments
This work was partially supported by the NASA grant NNX10AC55G, the International Space Science Institute (Bern) and Nordita (Stockholm). The authors thank Phil Goode, Vasyl Yurchshin, Valentyna Abramenko, and participants of the Nordita and ISSI teams for interesting discussions and useful suggestions.

\end{document}